\documentclass[10pt]{article}
\usepackage{multicol}
\usepackage{graphicx}
\usepackage{amsmath}
\usepackage[a4paper]{geometry}
\usepackage{enumerate}
\usepackage{hyperref}

\begin{document}

\begin{center}
{\Large \bf Dependence of temperatures and kinetic freeze-out
volume on centrality in Au-Au and Pb-Pb collisions at high energy}

\vskip1.0cm

Muhammad Waqas$^{1,}${\footnote{E-mail: waqas\_phy313@yahoo.com}},
Fu-Hu Liu$^{1,}${\footnote{Corresponding author. E-mail:
fuhuliu@163.com; fuhuliu@sxu.edu.cn}}, Zafar
Wazir$^{2,}${\footnote{E-mail: zafar.wazir@iiu.edu.pk}}
\\

{\small\it $^1$Institute of Theoretical Physics \& State Key
Laboratory of Quantum Optics and Quantum Optics Devices,\\ Shanxi
University, Taiyuan, Shanxi 030006, China

$^2$Department of Physics, International Islamic University, H-10,
Islamabad, Pakistan}

\end{center}

\vskip1.0cm

{\bf Abstract:} Centrality-dependent double-differential
transverse momentum spectra of negatively charged particles
($\pi^-$, $K^-$ and $\bar p$) at mid-(pseudo)rapidity interval in
nuclear collisions are analyzed by the standard distribution in
terms of multi-component. The experimental data measured in
gold-gold (Au-Au) collisions by the PHENIX Collaboration at the
Relativistic Heavy Ion Collider (RHIC) and in lead-lead (Pb-Pb)
collisions by the ALICE Collaboration at the Large Hadron Collider
(LHC) are studied. The effective temperature, initial temperature,
kinetic freeze-out temperature, transverse flow velocity and
kinetic freeze-out volume are extracted from the fitting to
transverse momentum spectra. We observed, that the mentioned five
quantities increase with the increase of event centrality due to
the fact that the average transverse momentum increases with the
increase of event centrality. This renders that larger momentum
(energy) transfer and further multiple-scattering had happened in
central centrality.
\\



\vskip1.0cm
\begin{multicols}{2}

{\section{Introduction}}

One of the most important questions in high energy collisions is
the identification of various phases of dense matter. It is
expected to reach a deconfined state of matter (quarks and gluons)
at high energy or density. This state of matter is called
Quark-Gluon Plasma (QGP), which was obtained in the early universe
shortly after the big-bang prior to the condensation in hadrons.
The characterization of phase transition in finite system is a
fascinating multi-disciplinary topic which has been studied for
decades~\cite{1,2} within different phenomenological applications.
The Relativistic Heavy Ion Collider (RHIC) and Large Hadron
Collider (LHC) have been providing excellent tools to determine
the phase structure of the strongly interacting Quantum
Chromodynamics (QCD) matter~\cite{3,4,5} and to study the
properties of QGP~\cite{6,7,8}.

Within the framework of statistical thermal models, it is assumed
that the initial stage of collisions of nuclei at the RHIC and
LHC~\cite{9,10,11} gives a tremendous amount of temperature, where
a hot and dense ``fireball" over an extended region for a very
short period of time (almost $10^{-22}$ seconds) is formed. The
fireball consists of QGP and it instantly cools which results in
the expansion of the system (the change of the volume or density
of the system) and the partons recombine into blizzard of hadronic
matter~\cite{12}. After the hadronization of the fireball, the
hadrons continuously interact with each other and the particle
number changes. This process results in decrease of temperature
and at a certain value where the reaction process stops and the
temperature at this point is called the ``chemical freeze-out
temperature" ($T_{ch}$). At the stage of chemical freeze-out, the
yield ratios of different types of particles remain
invariant~\cite{13}.

However, the rescattering process still take place which continues
to build up the collective (hydrodynamical) expansion.
Resultantly, the matter becomes dilute and the mean free path of
the given hadrons for the elastic reaction processes become
comparable with the size of the system. At this stage, the
rescattering process stops, which results in the decoupling of
hadrons from the rest of the system~\cite{14}. This stage is
called as the kinetic or thermal freeze-out stage, and the
temperature at this stage is called the kinetic or thermal
freeze-out temperature ($T_0$). After this stage the particle's
energy/momentum spectrum is frozen in time and it is the least
stage of the system evolution. Meanwhile, at this stage,
particle's movement is also affected by the flow effect which
should be excluded when one extracts $T_0$. To describe the flow
effect, one may use the transverse flow velocity $\beta_T$.

The above discussed $T_0$ and $\beta_T$ can be extracted from
transverse momentum ($p_T$) spectra of particles. Also, from $p_T$
spectra, one can extract the initial temperature $T_i$ according
to the color string percolation model~\cite{45,46,47}. Generally,
if the flow effect is not excluded in the temperature parameter,
this type of temperature is called the effective temperature
($T$). At least, three types of temperatures, $T$, $T_i$ and
$T_0$, can be extracted from $p_T$ spectra. Although the yield
ratios of different types of particles can be also obtained from
the normalizations of $p_T$ spectra and then $T_{ch}$ can be also
extracted from $p_T$ spectra, we mainly extract other three types
of temperatures and $\beta_T$ from $p_T$ spectra in this paper due
to their more pending situations.

In addition, volume is also an important parameter in high energy
collisions. The volume occupied by the ejectiles when the mutual
interactions become negligible, and the only force they feel is
the columbic repulsive force, is known as the kinetic freeze-out
volume ($V$). There are various freeze-out volumes at various
freeze-out stages, but we are only focusing on the kinetic
freeze-out volume $V$ in the present work. As we know, $V$ gives
the information of the co-existence of phase-transition, and is
important in the extraction of multiplicity, micro-canonical heat
capacity and it's negative branch or shape of the caloric curves
under the thermal constraints~\cite{15,16,17,18,19}. In this
paper, the fifth quantity extracted from $p_T$ spectra is $V$. By
way of parenthesis, the mean $p_T$, i.e. $\langle p_T\rangle$, is
also obtained.

The study of three types of temperatures, transverse flow velocity
and kinetic freeze-out volume is very wide, interesting and of
course a huge project. However, in this paper we will only analyze
the centrality dependences of the five quantities in gold-gold
(Au-Au) collisions at 200 GeV and in lead-lead (Pb-Pb) collisions
at 2.76 TeV. Only the $p_T$ spectra of negatively charged pions
($\pi^-$) and kaons ($K^-$) and antiprotons ($\bar p$) are used in
the extraction. These representational spectra are enough to
extract the necessary centrality dependences.

The remainder of this paper orderly consists of the method and
formalism, results and discussion, and summary and conclusions
which are presented in section 2, 3 and 4 respectively.
\\

{\section{The method and formalism}}

Soft excitation and hard scattering processes are the two
generally considered processes for the particle production. Soft
excitation process contributes in a narrow $p_T$ range which is
less than 2--3 GeV/$c$ or up to 4--5 GeV/$c$ and is responsible
for the production of most of the light flavored particles. The
soft excitation process has various choices of formalisms
including but are not limited to the Hagedorn thermal model
(Statistical-Bootstrap model)~\cite{20}, the standard
distribution~\cite{21}, the blast-wave model with Boltzmann-Gibbs
statistics~\cite{22,23,24}, the blast-wave model with Tsallis
statistics~\cite{25,26,27}, and current thermodynamical related
models~\cite{28,29,30,31}. The main contributor to the produced
particles is the soft excitation process.

If necessary, for the hard excitation process, there is limited
choice of formalisms~\cite{32,33,34} and can be described by the
theory of strong interaction. In fact, the contribution of hard
scattering process is parameterized to an inverse power law, i.e.
the Hagedorn function~\cite{20}
\begin{align}
f_H(p_T)=Ap_T\bigg(1+\frac{p_T}{p_0}\bigg)^{-n},
\end{align}
where $p_0$ and $n$ are free parameters, and $A$ is the normalized
constant related to the free parameters. The inverse power law is
obtained from the calculus of QCD~\cite{3,4,5}, and has at least
three revisions, which is out of focus of the present work and
will not be discussed further.

Different probability density functions can be used to describe
the contributions of soft excitation and hard scattering
processes. Due to few fraction and being earlier than the kinetic
freeze-out stage, the hard scattering process does not contribute
largely to $T_0$ and $\beta_T$ in general. In fact, the
contribution of hard scattering process can be neglected if we
study the $p_T$ spectra in a narrow range, say $p_T<2$--3 GeV/$c$
or extending to $p_T<4$--5 GeV/$c$, for which only the
contribution of soft excitation process is indeed needed. In our
opinion, various distributions show similar behaviors in case of
fitting the data with acceptable representations, which results in
similar $\langle p_T\rangle$ ($\sqrt{\langle p_T^{2}\rangle}$)
with different parameters.

For the spectra contributed by the soft excitation process, we can
choose the standard distribution, as it is very close in concept
to the ideal gas model. The standard distribution is the
combination of Boltzmann, Fermi-Dirac and Bose-Einstein
distributions. The probability density function of the standard
distribution in terms of $p_T$ at mid-rapidity is generally
as~\cite{21}:
\begin{align}
f_S(p_T)&=C\frac{gV'}{(2\pi)^2}{p_T}\sqrt{p_T^2+m_0^2} \nonumber\\
&\times\bigg[\exp\bigg(\frac{\sqrt{p_T^2+m_0^2}}{T}\bigg)+S\bigg]^{-1},
\end{align}
where $C$ is the normalization constant, $V'$ is the fitted
kinetic freeze-out volume (in terms of interaction volume) of the
emission source at the stage of kinetic freeze-out as discussed
above, $g$ is the degeneracy factor for pion and kaon (or proton)
and has the value of 3 (or 2), $m_0$ is the rest mass of the
considered particle, $S=-1$ (+1) is for bosons (fermions), and $T$
is the effective temperature as discussed above.

By considering the experimental rapidity range [$y_{\min}$,
$y_{\max}$] around the mid-rapidity, the accurate form of Eq. (2)
is~\cite{21}
\begin{align}
f_S(p_T)&= C\frac{gV'}{(2\pi)^2}p_T\sqrt{p_T^2+m_0^2}
\int_{y_{\min}}^{y_{\max}} \cosh y \nonumber\\
&\times\bigg[\exp\bigg(\frac{\sqrt{p_T^2+m_0^2}\cosh
y}{T}\bigg)+S\bigg]^{-1}dy.
\end{align}
$T$ and $V'$ in Eqs.(2) and (3) are free parameters. In most
cases, the single component standard distribution is not enough
for the simultaneous description of low-$p_T$ region. So we have
to use a two-component standard distribution.

In particular, in some cases, the multi-component ($l$-component)
standard distribution has to be used, which can be demonstrated
as:
\begin{align}
f_S(p_T)&=\sum\limits_{i=1}^{l}k_i C_i
\frac{gV'_i}{(2\pi)^2}p_T\sqrt{p_T^2+m_0^2}
\int_{y_{\min}}^{y_{\max}} \cosh y \nonumber\\
&\times\bigg[\exp\bigg(\frac{\sqrt{p_T^2+m_0^2}\cosh
y}{T_i}\bigg)+S\bigg]^{-1}dy,
\end{align}
where $C_i$ is the normalization constant, $k_i$ represents the
fraction contributed by the $i$th component, and $T_i$ and $V'_i$
are free parameter denoted the kinetic freeze-out temperature and
volume respectively corresponding to the $i$th component.

In case of considering both the contributions of soft excitation
and hard scattering processes, we use the superposition in
principle
\begin{align}
f_0({p_T})=k_Sf_S(p_T)+(1-k_S)f_H(p_T),
\end{align}
where $k_S$ is the contribution ratio of soft excitation process.
The contribution ranges of the soft excitation and hard scattering
processes described by Eq. (5) are intersectant in low-$p_T$
region.

Another type of superposition which uses the usual step function
$\theta(x)$ based on the Hagedorn model~\cite{20} is
\begin{align}
f_0(p_T)=A_1{\theta}(p_1-p_T)f_S(p_T)+
A_2{\theta}(p_T-p_1)f_H(p_T),
\end{align}
where $A_1$ and $A_2$ are the normalized constants which
synthesize ${A_1}{f_S}(p_1)={A_2}{f_H}(p_1)$. The contribution
ranges of the soft excitation and hard scattering processes
described by Eq. (6) are segregative at $p_T=p_1$.

In the present work, we will study only the first component in
Eqs. (5) or (6), because we do not study a very wide $p_T$ range.
Meanwhile, we use the two-component standard distribution, i.e.
$l=2$ in Eq. (4) in which the first or second components has no
particular priority. As probability density functions, Eqs.
(1)--(6) are normalized to 1 naturally. When we compare the
probability density functions with the experimental data which
appear usually in other forms, a suitable transformation for the
probability density function is certainly needed. Considering the
treatment of normalization, the real fitted kinetic freeze-out
volume should be $V'_i/(C_ik_i)$ which will be simply used in the
following section as the symbols, $V_i$ or $V$ ($=\sum_{i=1}^l
V_i$).
\\

{\section{Results and discussion}}

Figure 1 presents the event centrality dependent double
differential $p_T$ spectra, $(1/2\pi p_T)d^2N/dp_Tdy$, of the
identified particles ($\pi^-$, $K^-$ and $\bar p$) produced in
Au-Au collisions at $\sqrt{s_{NN}}=200$ GeV in the
mid-pseudorapidity interval of $|\eta|<0.35$, where $N$ and $y$
stands for the number of particles and rapidity respectively. The
symbols are the the experimental data measured by the PHENIX
Collaboration~\cite{35}. The spectra of centralities 0--5\%,
5\%--10\%, 10\%--15\%, 15--20\%, 20--30\%, 30--40\%, 40--50\%,
50-60\%, 60--70\%, 70--80\% and 80--92\% in the three panels are
multiplied by 20, 10, 5, 2.5, 1.5, 1, 1, 1, 1, 1 and 1,
respectively. In addition, the spectra of centralities 15--20\%,
20--30\%, 30--40\%, 40--50\%, 50-60\%, 60--70\%, 70--80\% and
80--92\% in panel (a) are re-multiplied by 0.8, 0.7, 0.6, 0.45,
0.3, 0.24, 0.17 and 0.15 respectively. The curves are our fitting
results by using the two-component standard distribution, Eq. (4)
with $l=2$. The substantially togethered event centralities, the
values of free parameters ($T_1$, $T_2$, $V_1$ and $V_2$),
$\chi^2$ and the number of degree of freedom (ndof) are listed in
Table 1. One can see the well approximate description of the model
results to the experimental data of the PHENIX Collaboration in
special $p_T$ ranges in high energy Au-Au collisions at the RHIC.

\begin{figure*}[htb!]
\begin{center}
\includegraphics[width=15cm]{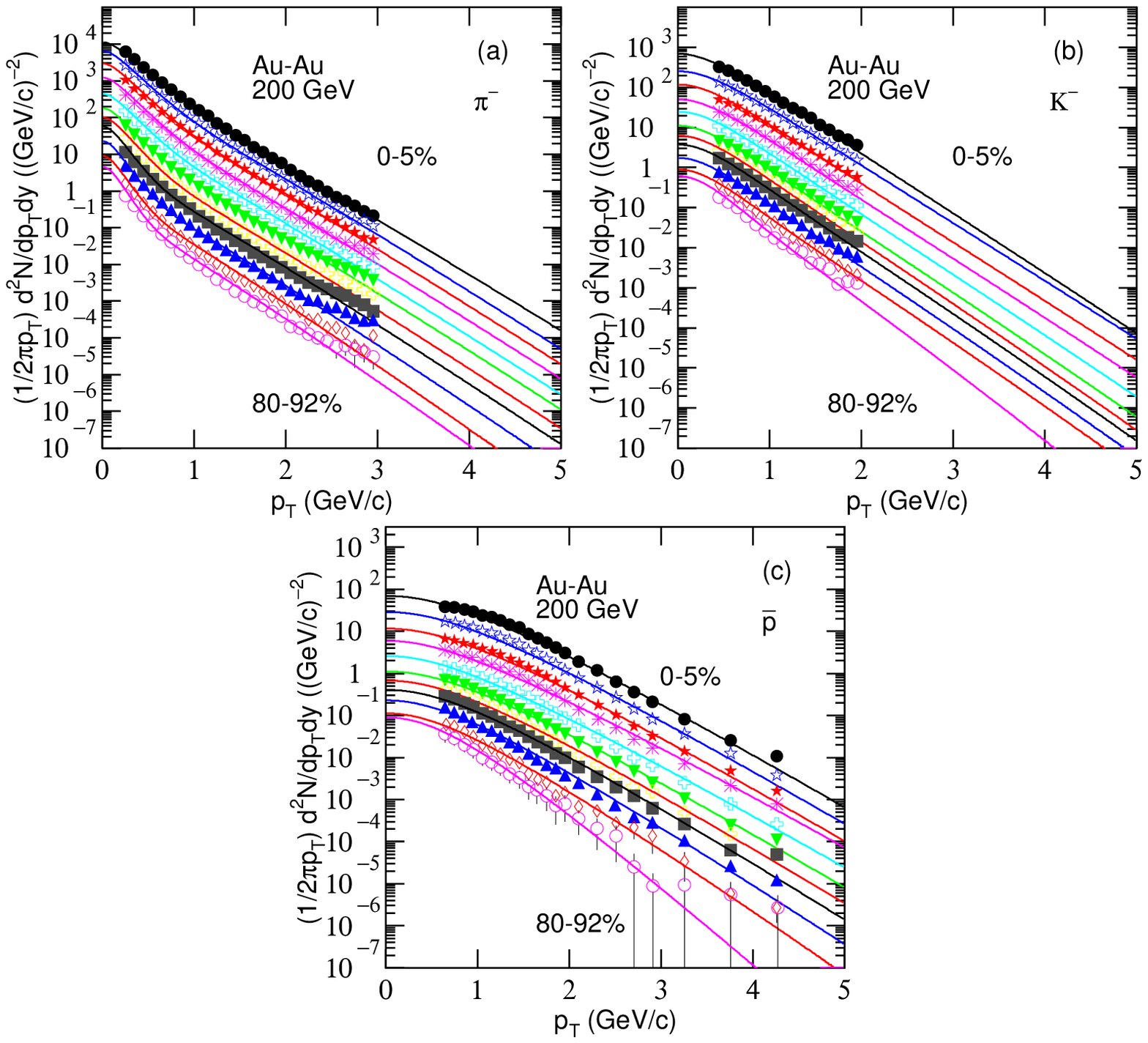}
\end{center}
Fig. 1. Centrality dependant $(1/2\pi p_T)d^2N/dp_Tdy$ of (a)
$\pi^-$, (b) $K^-$ and (c) $\bar p$ produced in $|\eta|<0.35$ in
Au-Au collisions at 200 GeV. The symbols represent experimental
data of the PHENIX Collaboration~\cite{35}, while the curves are
the results of our fits by using the two-component standard
distribution, Eq. (4) with $l=2$. The spectra of centralities
0--5\%, 5\%--10\%, 10\%--15\%, 15--20\%, 20--30\%, 30--40\%,
40--50\%, 50-60\%, 60--70\%, 70--80\% and 80--92\% in the three
panels are multiplied by 20, 10, 5, 2.5, 1.5, 1, 1, 1, 1, 1 and 1,
respectively. In addition, the spectra of centralities 15--20\%,
20--30\%, 30--40\%, 40--50\%, 50-60\%, 60--70\%, 70--80\% and
80--92\% in panel (a) are re-multiplied by 0.8, 0.7, 0.6, 0.45,
0.3, 0.24, 0.17 and 0.15 respectively.
\end{figure*}

\begin{figure*}[htb!]
\begin{center}
\includegraphics[width=15cm]{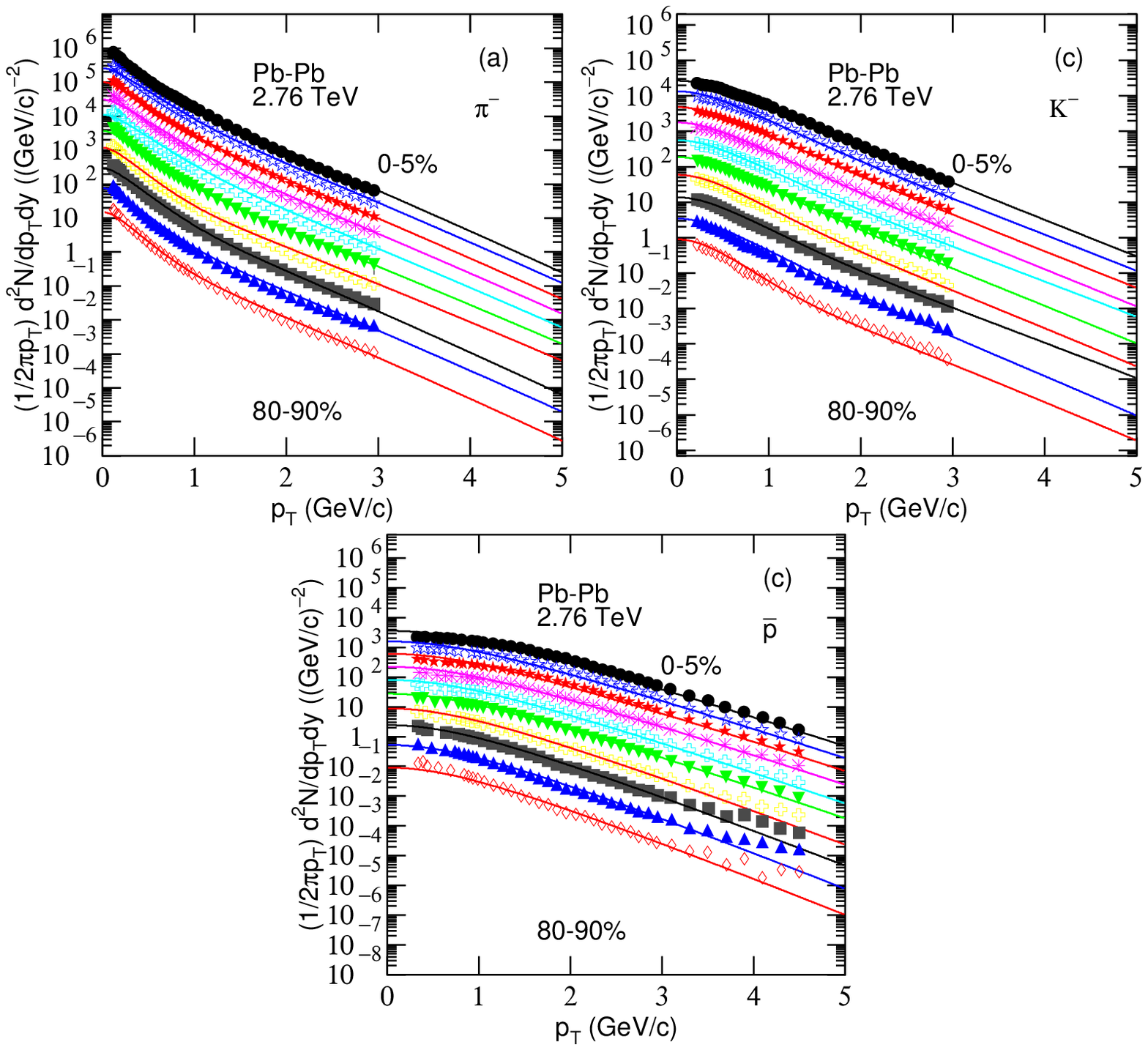}
\end{center}
Fig. 2. Same as Fig. 1, but for the spectra of $\pi^-$, $K^-$ and
$\bar p$ in $|y|<0.5$ in Pb-Pb collisions at $\sqrt{s_{NN}}=2.76$
TeV. The symbols represent the measured data of the ALICE
Collaboration~\cite{36}, where the spectra are scaled by factors
$2^n$ and $n$ changes from 9 to 0 as the event centrality changes
from 0--5\% to 80--90\%.
\end{figure*}

\begin{table*}
{\scriptsize Table 1. Values of effective temperatures ($T_1$ and
$T_2$), volumes ($V_1$ and $V_2$), $\chi^2$, and number of degree
of freedom (ndof) corresponding to the curves in Figs. 1--2.
\vspace{-.50cm}
\begin{center}
\begin{tabular}{ccccccccc}\\ \hline\hline
Figure & Particle & Centrality& $T_1$ (GeV) & $T_2$ (GeV)& $V_1$
((fm)$^3$) & $V_2$ ((fm)$^3$) & $\chi^2$ & ndof
\\\hline
Fig. 1  & $\pi^-$  & 0--5\%   & $0.181\pm0.014$ & $0.268\pm0.006$ & $745\pm100$ & $1967\pm309$ & 45  & 24\\
Au-Au   &          & 5--10\%  & $0.138\pm0.013$ & $0.270\pm0.006$ & $564\pm80$  & $2054\pm174$ & 14  & 24\\
200 GeV &          & 10-15\%  & $0.179\pm0.015$ & $0.250\pm0.005$ & $430\pm62$  & $2004\pm153$ & 233 & 24\\
        &          & 15--20\% & $0.167\pm0.012$ & $0.247\pm0.003$ & $354\pm61$  & $1862\pm320$ & 63  & 24\\
        &          & 20--30\% & $0.151\pm0.009$ & $0.245\pm0.004$ & $360\pm39$  & $1854\pm216$ & 87  & 24\\
        &          & 30--40\% & $0.125\pm0.015$ & $0.237\pm0.006$ & $195\pm30$  & $2005\pm189$ & 276 & 24\\
        &          & 40--50\% & $0.148\pm0.014$ & $0.227\pm0.004$ & $106\pm25$  & $1585\pm171$ & 388 & 24\\
        &          & 50--60\% & $0.098\pm0.011$ & $0.227\pm0.003$ & $59\pm11$   & $1048\pm153$ & 146 & 24\\
        &          & 60--70\% & $0.125\pm0.010$ & $0.274\pm0.006$ & $37\pm11$   & $984\pm109$  & 71  & 24\\
        &          & 70--80\% & $0.112\pm0.014$ & $0.209\pm0.007$ & $18\pm3$    & $296\pm37$   & 256 & 24\\
        &          & 80-92\%  & $0.068\pm0.015$ & $0.221\pm0.006$ & $8.0\pm1.0$ & $213\pm13$   & 69  & 24\\
\cline{2-9}
        & $K^-$    & 0--5\%   & $0.201\pm0.009$ & $0.271\pm0.008$ & $196\pm36$      & $1081\pm178$ & 17 & 12\\
        &          & 5--10\%  & $0.188\pm0.007$ & $0.255\pm0.008$ & $32\pm6$        & $1116\pm71$  & 18 & 12\\
        &          & 10--15\% & $0.211\pm0.013$ & $0.250\pm0.006$ & $37\pm6$        & $1030\pm102$ & 24 & 12\\
        &          & 15--20\% & $0.135\pm0.014$ & $0.248\pm0.008$ & $35\pm5$        & $836\pm105$  & 14 & 12\\
        &          & 20--30\% & $0.219\pm0.015$ & $0.241\pm0.006$ & $10\pm1$        & $733\pm66$   & 18 & 12\\
        &          & 30--40\% & $0.156\pm0.010$ & $0.238\pm0.007$ & $4.0\pm0.3$     & $665\pm82$   & 13 & 12\\
        &          & 40--50\% & $0.130\pm0.011$ & $0.235\pm0.008$ & $13\pm1.2$      & $253\pm39$   & 24 & 12\\
        &          & 50--60\% & $0.170\pm0.013$ & $0.230\pm0.006$ & $1.6\pm0.2$     & $577\pm48$   & 9  & 12\\
        &          & 60--70\% & $0.163\pm0.015$ & $0.225\pm0.006$ & $0.50\pm0.02$   & $544\pm50$   & 35 & 12\\
        &          & 70--80\% & $0.155\pm0.015$ & $0.220\pm0.007$ & $0.30\pm0.01$   & $505\pm60$   & 215& 12\\
        &          & 80--92\% & $0.130\pm0.010$ & $0.218\pm0.005$ & $0.030\pm0.001$ & $460\pm130$  & 66 & 12\\
\cline{2-9}
        & $\bar p$ & 0--5\%   & $0.280\pm0.017$ & $0.320\pm0.003$ & $36\pm7$      & $786\pm120$ & 17  & 18\\
        &          & 5--10\%  & $0.272\pm0.015$ & $0.317\pm0.002$ & $31\pm7$      & $775\pm70$  & 20  & 18\\
        &          & 10-15\%  & $0.265\pm0.018$ & $0.313\pm0.006$ & $16\pm3$      & $747\pm86$  & 162 & 18\\
        &          & 15-20\%  & $0.185\pm0.016$ & $0.313\pm0.005$ & $23\pm4$      & $710\pm80$  & 35  & 18\\
        &          & 20--30\% & $0.211\pm0.013$ & $0.309\pm0.006$ & $13\pm1$      & $678\pm65$  & 41  & 18\\
        &          & 30-40\%  & $0.247\pm0.018$ & $0.317\pm0.007$ & $10\pm1$      & $634\pm50$  & 11  & 18\\
        &          & 40--50\% & $0.244\pm0.015$ & $0.308\pm0.008$ & $4.0\pm0.5$   & $597\pm37$  & 39  & 18\\
        &          & 50--60\% & $0.237\pm0.013$ & $0.300\pm0.004$ & $1.6\pm0.3$   & $574\pm70$  & 43  & 18\\
        &          & 60--70\% & $0.210\pm0.013$ & $0.285\pm0.006$ & $0.50\pm0.02$ & $476\pm45$  & 28  & 18\\
        &          & 70--80\% & $0.170\pm0.014$ & $0.267\pm0.005$ & $0.30\pm0.01$ & $455\pm50$  & 110 & 18\\
        &          & 80--92\% & $0.172\pm0.010$ & $0.218\pm0.004$ & $0.40\pm0.01$ & $406\pm40$  & 5   & 18\\
\hline
Fig. 2  & $\pi^-$  & 0--5\%   & $0.179\pm0.010$ & $0.340\pm0.004$ & $2517\pm208$ & $4099\pm252$ & 206 & 37\\
Pb-Pb   &          & 5--10\%  & $0.172\pm0.012$ & $0.331\pm0.004$ & $2226\pm135$ & $3468\pm201$ & 101 & 37\\
2.76 TeV&          & 10-20\%  & $0.165\pm0.010$ & $0.312\pm0.005$ & $1395\pm70$  & $2813\pm176$ & 70  & 37\\
        &          & 20--30\% & $0.161\pm0.014$ & $0.300\pm0.006$ & $1012\pm103$ & $2427\pm231$ & 94  & 37\\
        &          & 30--40\% & $0.147\pm0.013$ & $0.280\pm0.005$ & $534\pm56$   & $3502\pm165$ & 244 & 37\\
        &          & 40--50\% & $0.129\pm0.012$ & $0.289\pm0.005$ & $307\pm38$   & $3600\pm198$ & 110 & 37\\
        &          & 50--60\% & $0.108\pm0.010$ & $0.281\pm0.005$ & $176\pm32$   & $3380\pm213$ & 206 & 37\\
        &          & 60--70\% & $0.114\pm0.015$ & $0.290\pm0.005$ & $94\pm13$    & $3150\pm147$ & 125 & 35\\
        &          & 70--80\% & $0.130\pm0.010$ & $0.287\pm0.007$ & $45\pm9$     & $2930\pm206$ & 502 & 35\\
        &          & 80--90\% & $0.118\pm0.014$ & $0.287\pm0.006$ & $16\pm2$     & $1945\pm102$ & 119 & 35\\
\cline{2-9}
        & $K^-$    & 0--5\%   & $0.277\pm0.010$ & $0.361\pm0.007$ & $170\pm36$    & $3442\pm208$ & 3   & 32\\
        &          & 5--10\%  & $0.270\pm0.009$ & $0.359\pm0.009$ & $152\pm10$    & $3237\pm240$ & 15  & 32\\
        &          & 10--20\% & $0.252\pm0.009$ & $0.354\pm0.007$ & $97\pm11$     & $3190\pm187$ & 8   & 32\\
        &          & 20--30\% & $0.265\pm0.009$ & $0.350\pm0.006$ & $53\pm7$      & $3136\pm200$ & 22  & 32\\
        &          & 30--40\% & $0.247\pm0.008$ & $0.395\pm0.006$ & $27\pm4$      & $3014\pm198$ & 12  & 32\\
        &          & 40--50\% & $0.210\pm0.008$ & $0.341\pm0.006$ & $22\pm4$      & $2828\pm190$ & 119 & 32\\
        &          & 50--60\% & $0.213\pm0.011$ & $0.345\pm0.005$ & $8.1\pm1.0$   & $2659\pm210$ & 26  & 32\\
        &          & 60--70\% & $0.226\pm0.009$ & $0.335\pm0.005$ & $5.8\pm1.3$   & $2132\pm176$ & 46  & 32\\
        &          & 70--80\% & $0.225\pm0.010$ & $0.332\pm0.006$ & $2.6\pm0.3$   & $1736\pm134$ & 71  & 32\\
        &          & 80-90\%  & $0.170\pm0.009$ & $0.320\pm0.006$ & $0.50\pm0.05$ & $1150\pm120$ & 75  & 32\\
\cline{2-9}
        & $\bar p$ & 0--5\%   & $0.426\pm0.012$ & $0.421\pm0.007$ & $55\pm6$      & $1965\pm195$ & 49  & 33\\
        &          & 5--10\%  & $0.300\pm0.010$ & $0.425\pm0.005$ & $75\pm6$      & $1782\pm164$ & 373 & 33\\
        &          & 10--20\% & $0.427\pm0.010$ & $0.397\pm0.005$ & $50\pm5$      & $1748\pm140$ & 43  & 33\\
        &          & 20--30\% & $0.405\pm0.013$ & $0.395\pm0.006$ & $31\pm4$      & $1690\pm130$ & 52  & 33\\
        &          & 30--40\% & $0.400\pm0.012$ & $0.382\pm0.005$ & $29\pm4$      & $1534\pm158$ & 20  & 33\\
        &          & 40--50\% & $0.352\pm0.011$ & $0.383\pm0.006$ & $15\pm2$      & $401\pm135$  & 109 & 33\\
        &          & 50--60\% & $0.331\pm0.012$ & $0.380\pm0.005$ & $6.0\pm0.5$   & $1357\pm125$ & 67  & 33\\
        &          & 60--70\% & $0.310\pm0.012$ & $0.369\pm0.006$ & $8.7\pm1.0$   & $1069\pm90$  & 75  & 33\\
        &          & 70--80\% & $0.321\pm0.015$ & $0.333\pm0.004$ & $1.0\pm0.3$   & $978\pm50$   & 104 & 33\\
        &          & 80-90\%  & $0.310\pm0.014$ & $0.328\pm0.006$ & $0.70\pm0.04$ & $708\pm60$   & 122 & 33\\
\hline
\end{tabular}%
\end{center}}
\end{table*}

Figure 2 is the same as Fig. 1 but it gives the results for
$\pi^-$, $K^-$ and $\bar p$ in different centrality bins in Pb-Pb
collisions at 2.76 TeV in the mid-rapidity interval $|y|<0.5$. The
experimental data of the ALICE Collaboration is represented by the
symbols~\cite{36}, where as the spectra of Pb-Pb is scaled by the
factor of $2^n$ and $n$ changes from 9 to 0 with the change of
event centrality such as from 0--5\% to 80--90\%. The related
parameters and the existed centralities are listed together in
Table 1. One can see the well approximate description of the model
results to the experimental data of the ALICE Collaboration in
special $p_T$ ranges in high energy Pb-Pb collisions at the LHC.

\begin{figure*}[htb!]
\begin{center}
\includegraphics[width=15cm]{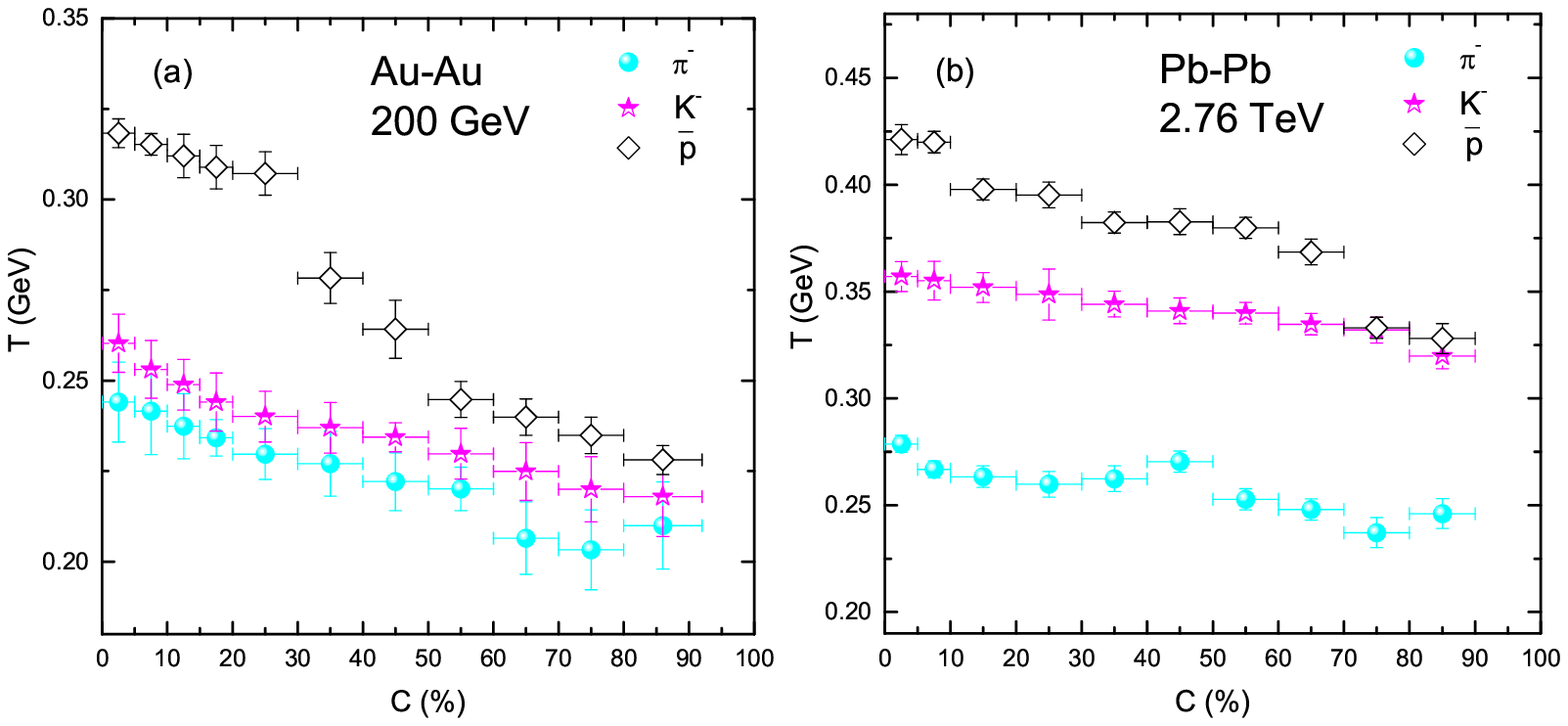}
\end{center}
Fig. 3. Dependence of effective temperature $T$ in different
centrality bins $C$ in (a) Au-Au collisions at 200 GeV and (b)
Pb-Pb collisions at 2.76 TeV. The obtained values corresponded to
identified particles are extracted from the experimental $p_T$
spectra.
\end{figure*}

To study the change trend of parameters with centrality, Fig. 3
shows the dependence of effective temperature on centrality for
the productions of $\pi^-$, $K^-$ and $\bar p$ in different
centrality bins in Au-Au and Pb-Pb collisions at 200 GeV and 2.76
TeV respectively. Panel (a) show the result for Au-Au collisions,
while panel (b) show the result for Pb-Pb collisions. Different
symbols represent different particles. One can see the clear
decrease of effective temperature from central to peripheral
collisions. The reason behind it is, the more violent collisions
in central collisions where it can get a degree of higher
excitation and also involve more number of participants in
interactions, while they decrease from central to peripheral
collisions. The effective temperatures in collisions at the LHC in
different centrality bins are higher than those at the RHIC due to
more energy deposition in collisions at the LHC.

\begin{figure*}[htb!]
\begin{center}
\vskip0.2cm
\includegraphics[width=15cm]{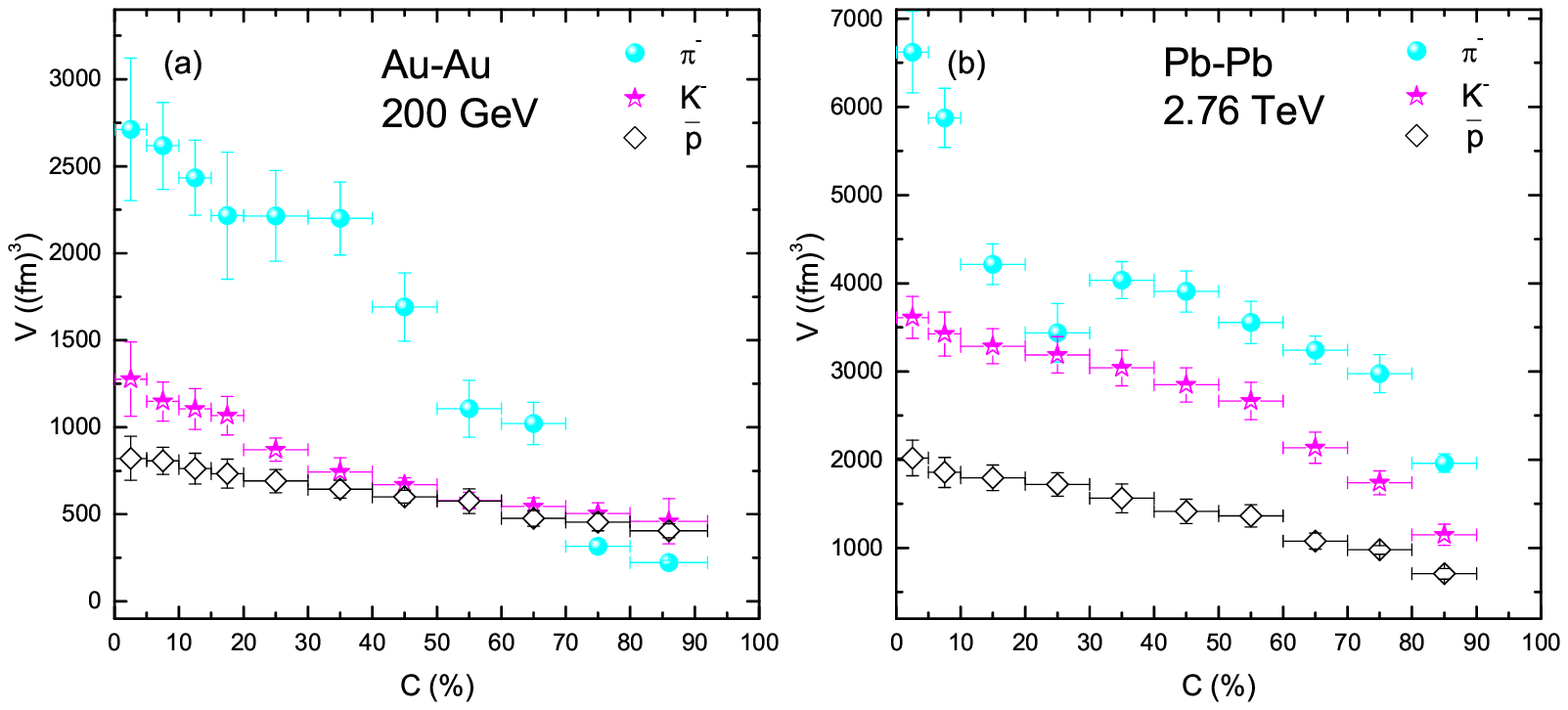}
\end{center}
Fig. 4. Same as Fig. 3, but for dependence of kinetic freeze-out
volume $V$ on centrality $C$.
\end{figure*}

Figure 4 is the same to Fig. 3, however, it shows the result for
the dependence of kinetic freeze-out volume $V$ in events with
different centralities $C$, where $V=V_1+V_2$ due to the
additivity of volume. One can see that the kinetic freeze-out
volume decreases from central to peripheral collisions, as the
number of participant nucleons decreases from central to
peripheral collisions depending on the interaction volume. Due to
large number of binary collisions by the rescattering of partons,
the system with more participants reaches quickly to equilibrium
state, but the decreases in centrality, the decreases the number
of participants and the system goes slowly to equilibrium state.
The large volume and more number of participants in the central
collisions may indicate the occupation of super-dense hadronic
matter, but of course further and more complete information about
the local energy density of super-hadronic matter is needed to
study the possible phase transition of QGP. The figure shows a
volume differential scenario too. The heavier the particle is, the
smaller the kinetic freeze-out volume has, which shows the early
freeze-out of heavier particles as compared to the lighter
particles and suggests different freeze-out surfaces for different
particles. Such result can be found in literature~\cite{39,40}.

\begin{figure*}[htb!]
\begin{center}
\vskip0.2cm
\includegraphics[width=15cm]{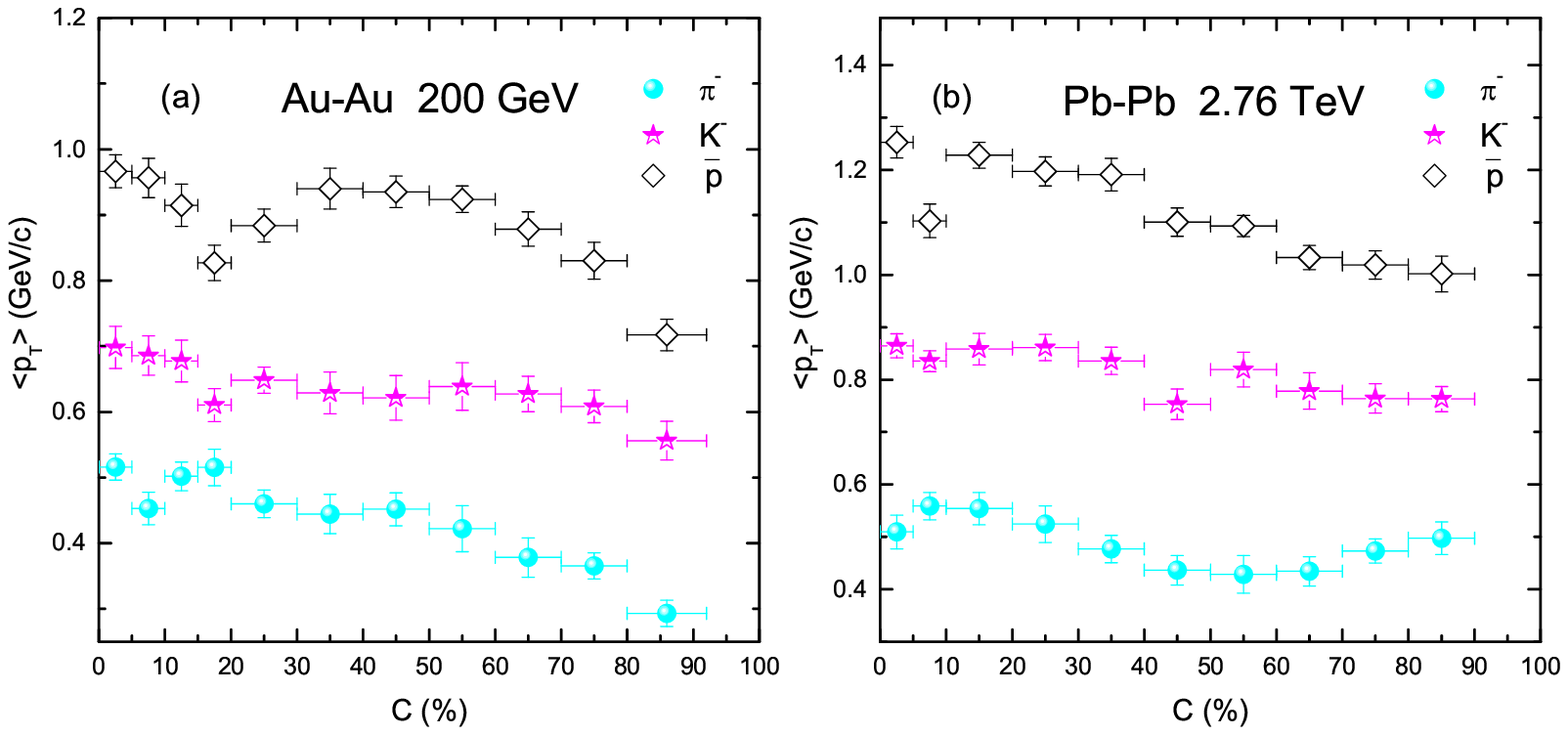}
\end{center}
Fig. 5. Same as Fig. 3, but for dependence of mean transverse
momentum $\langle p_T\rangle$ on centrality $C$.
\end{figure*}

\begin{figure*}[htb!]
\begin{center}
\vskip0.2cm
\includegraphics[width=15cm]{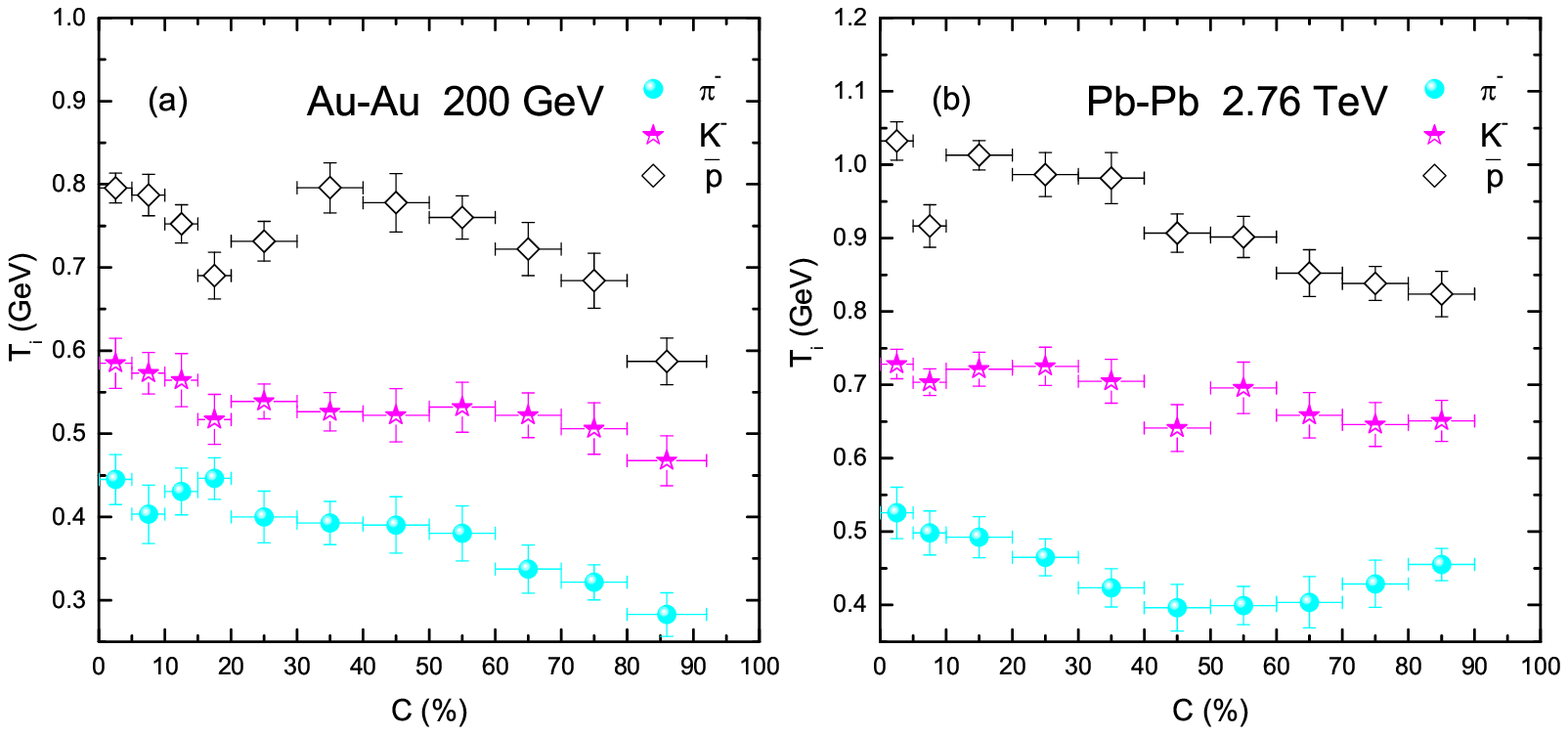}
\end{center}
Fig. 6. Same as Fig. 3, but for dependence of initial temperature
$T_i$ on centrality $C$.
\end{figure*}

The dependence of mean transverse momentum $\langle p_T\rangle$ in
different centrality events is shown in Fig. 5. The symbols
represent $\langle p_T\rangle$ for different particles obtained
from the fitting function Eq. (4) with $l=2$ over a $p_T$ range
from 0 to 5 GeV/$c$, where the parameter values are listed in
Table 1. One can see that $\langle p_T\rangle$ decreases from
central to peripheral Au-Au and Pb-Pb collisions for all particle
species, and it may be caused due to decreasing the participant
nucleons from central to peripheral collisions and this result is
similar to ref.~\cite{41}. It is also important to notice that
$\bar p$ spectra exhibit a concave shape in the peripheral events,
which is well described by the power law parametrization as
observed in ref.~\cite{42}, but this curvature decreases with the
increasing centrality and it leads to an almost exponential
dependence on $\langle p_T\rangle$ for the most central
collisions. Furthermore, $\langle p_T\rangle$ for heavier particle
is larger than that for lighter ones, and $\langle p_T\rangle$ at
LHC energy is slightly larger than that at RHIC energy. The
increase of $\langle p_T\rangle$ in central collisions and with
the massive mass of the particle may indicate the collective
radial flow, and the same behavior is observed at a few
GeV~\cite{43} and more than 10 GeV~\cite{44}.

Figure 6 is the same to Fig. 5, however it demonstrates the result
for the initial temperature $T_i$, where $T_i$ is obtained by the
root-mean-square $p_T$ divide by $\sqrt{2}$, i.e. $\sqrt{\langle
p_T^{2}\rangle/2}$ according to the color string percolation
model~\cite{45,46,47}. The symbols are the representation of the
results obtained from the fitting function Eq.(4) with $l=2$ over
a $p_T$ range in 0--5 GeV/$c$ and with the parameter values listed
in Table 1. The mass differential temperature scenario is also
observed. It is necessary to point out that the initial
temperature obtained in this work is larger than the effective
temperature which is in agreement with the order of time evolution
of interacting system.

\begin{figure*}[htb!]
\begin{center}
\vskip0.2cm
\includegraphics[width=15cm]{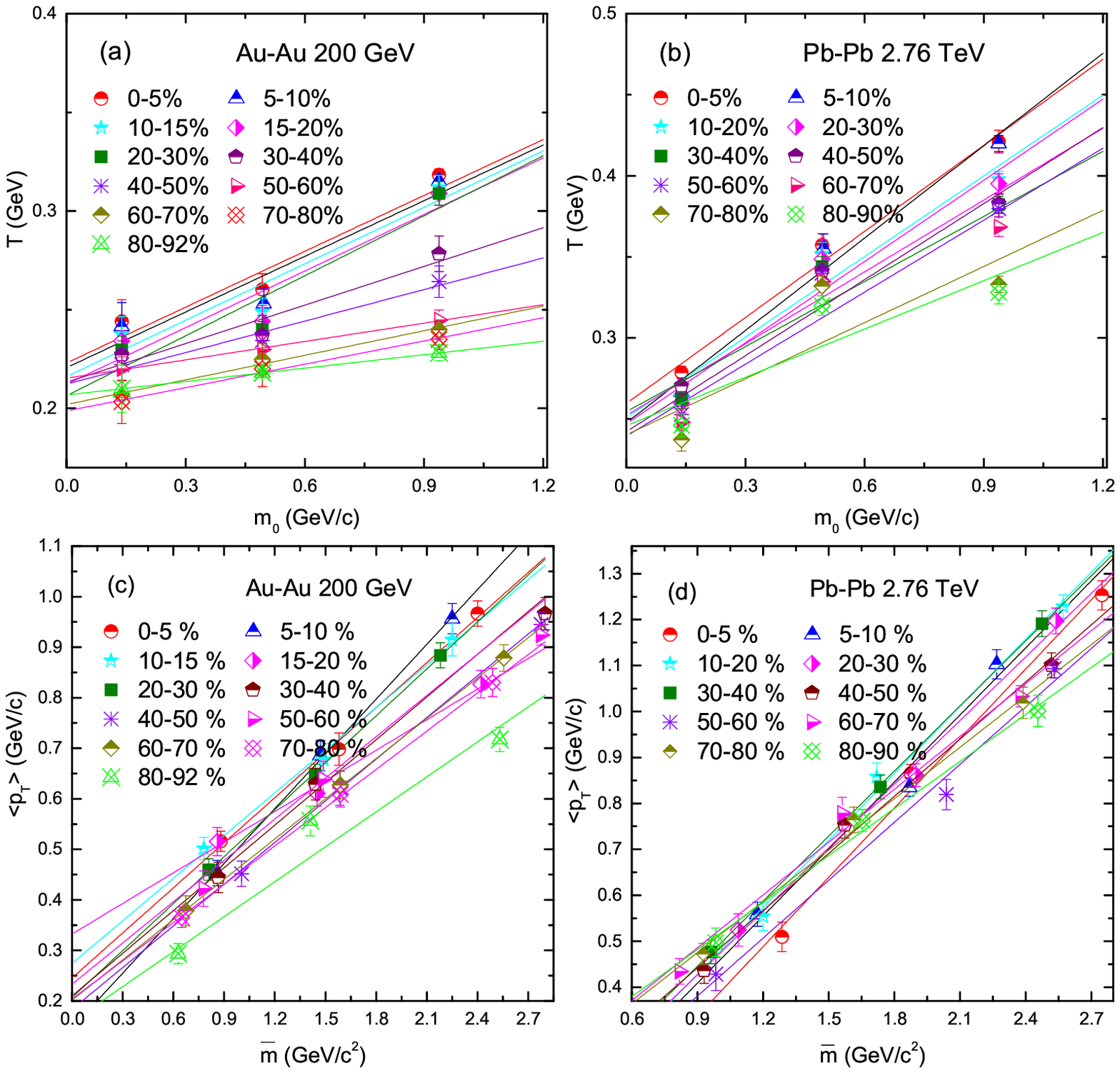}
\end{center}
Fig. 7. Dependence of (a)(b) effective temperature $T$ on rest
mass $m_0$ and (c)(d) mean transverse momentum $\langle
p_T\rangle$ on mean energy or mean moving mass $\overline{m}$ for
(a)(c) Au-Au collisions at 200 GeV and (b)(d) Pb-Pb collisions at
2.76 TeV in different centrality classes. The lines are the
results fitted for the values of derived quantities.
\end{figure*}

\begin{figure*}[htb!]
\begin{center}
\vskip1cm
\includegraphics[width=15cm]{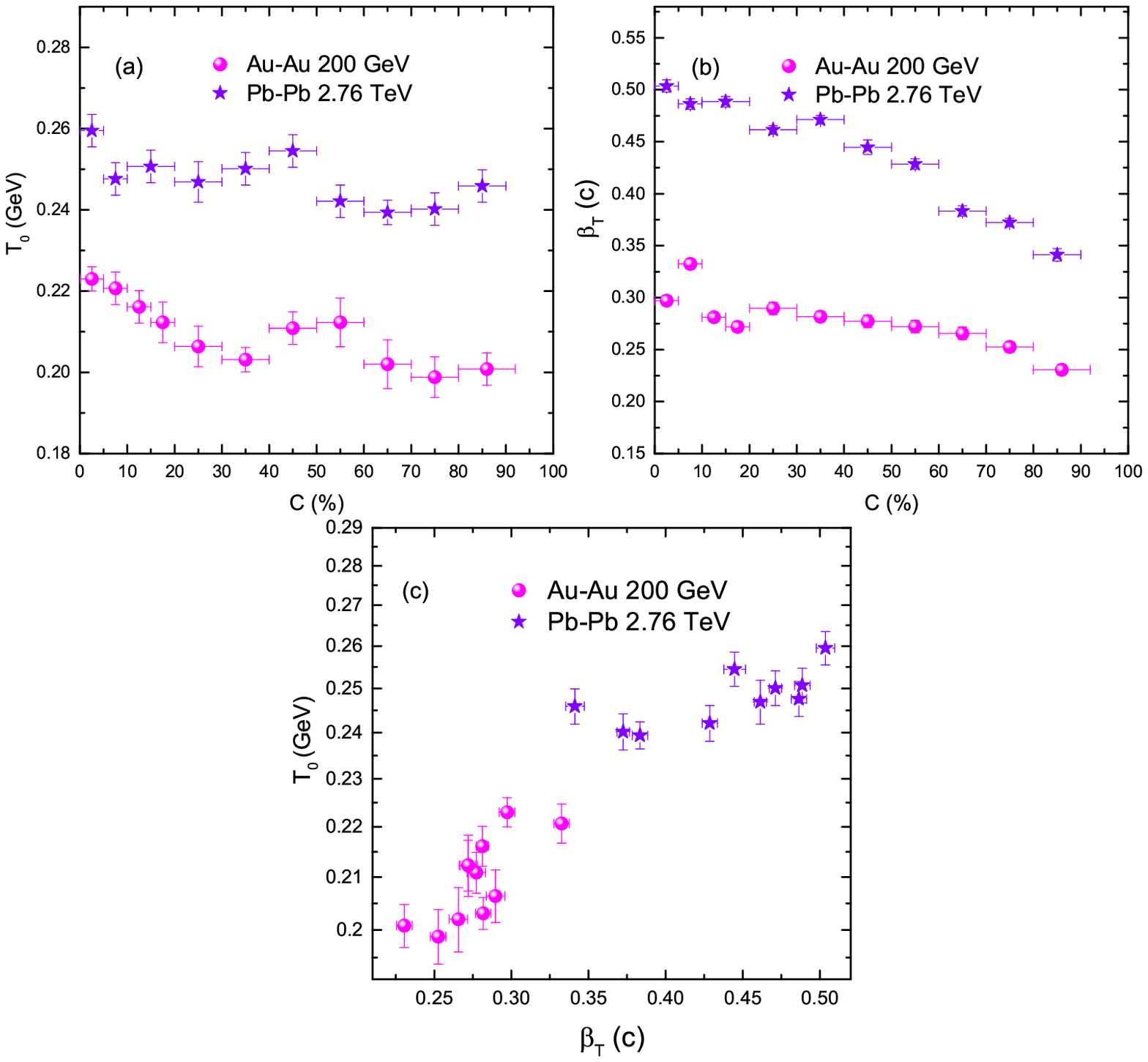}
\end{center}
Fig. 8. Dependence of (a) kinetic freeze-out temperature $T_0$ on
centrality $C$ and (b) transverse flow velocity $\beta_T$ on
centrality $C$.
\end{figure*}

\begin{figure*}[htb!]
\begin{center}
\vskip0.5cm
\includegraphics[width=7.5cm]{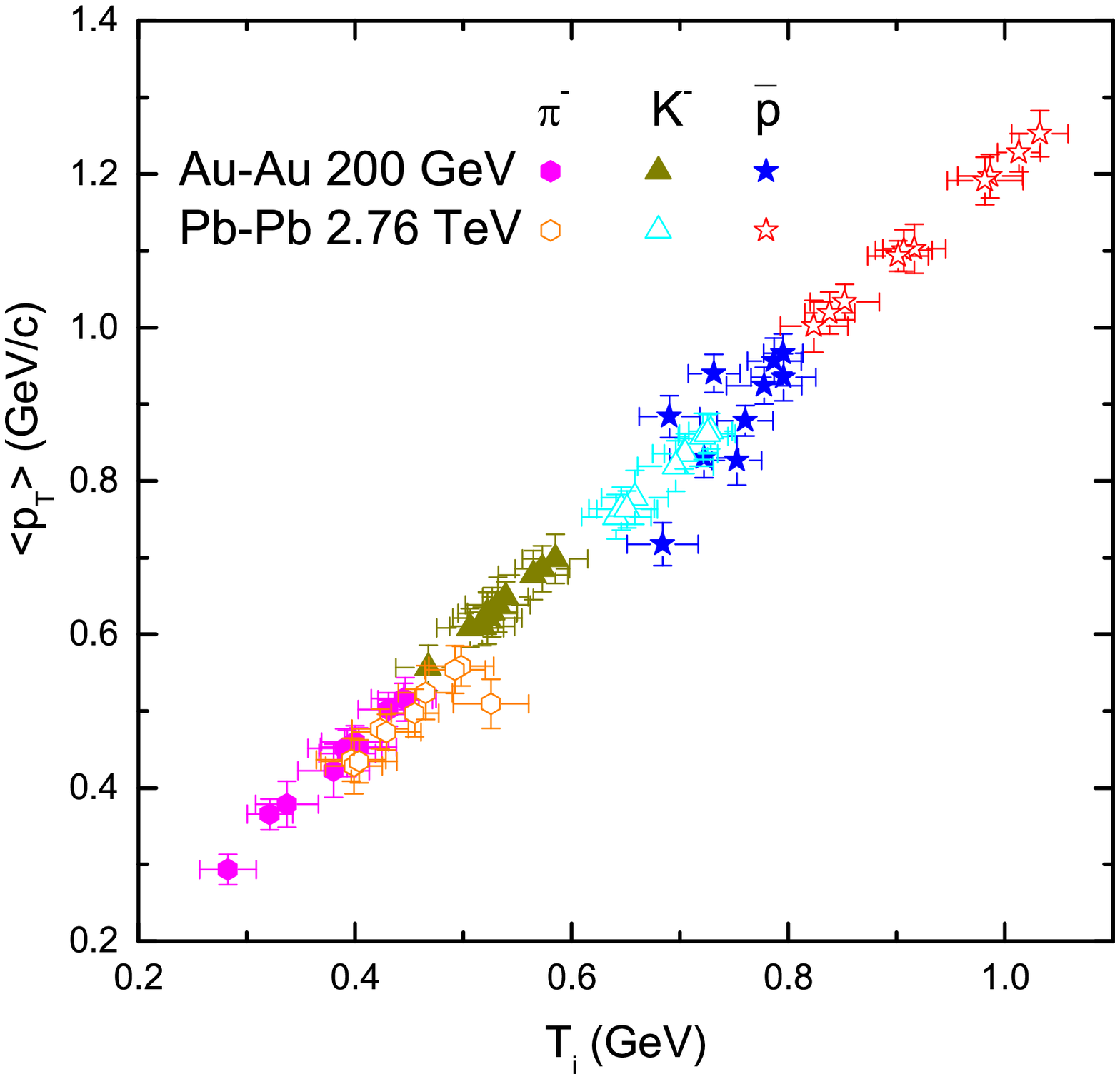}
\end{center}
Fig. 9. Dependence of mean transverse momentum $\langle
p_T\rangle$ on initial temperature $T_i$.
\end{figure*}

Figure 7 shows the dependence of (a)(b) effective temperature $T$
on rest mass $m_0$ and (c)(d) mean transverse momentum $\langle
p_T\rangle$ on mean energy or mean moving mass $\overline{m}$ in
the rest frame of emission source for (a)(c) Au-Au collisions at
200 GeV and (b)(d) Pb-Pb collisions at 2.76 TeV in different
centrality classes. The symbols represent the derived quantities
according to the free parameters listed in Table 1, where
$T=(T_1V_1+T_2V_2)/(V_1+V_2)$, $\langle p_T\rangle=\int
p_Tf_S(p_T)dp_T$, and
$\overline{m}=\langle\sqrt{p_T^2+p_z^2+m_0^2}\rangle$, where $p_z$
is longitudinal momentum and its distribution can be obtained
according to $p_T$ distribution if we assume isotropic emission in
the source rest frame~\cite{44a}. The lines are the results fitted
for the values of derived quantities if we assume linear
correlations are existent.

The intercepts in Figs. 7(a) and 7(b) are regarded as the kinetic
freeze-out temperature $T_0$, and the slopes in Figs. 7(c) and
7(d) are regarded as the transverse flow velocity
$\beta_T$~\cite{44b,44c,44d,44e}. The dependences of (a) $T_0$ on
$C$ and (b) $\beta_T$ on $C$, as well as the correlation between
(c) $\beta_T$ and $\beta_T$ are presented in Fig. 8. One can see
that $T_0$ and $\beta_T$ decrease with the increase of $C$, $T_0$
increases with the increase of $\beta_T$, which renders that
central collisions display higher excitation and quicker expansion
than peripheral collisions due to more energy deposition in
central collisions.

It should be noted that $T_0$ shown in Fig. 8 is larger than
160--170 MeV which is regarded as the chemical freeze-out
temperature $T_{ch}$ of phase transition from hadronic matter to
QGP. As expectation, $T_0$ should be less than or equal to
$T_{ch}$ due to time evolution. The difference between our results
and $T_{ch}$ is regarded as different ``thermometers" (methods)
used in the extraction of temperature. In our opinion, an unified
``thermometer" (method) should be used in the extraction of
temperature. Or, one may find a relation to convert one
temperature to another one, as what one did between Celcius
Temperature and degree Fahrenheit.

Meanwhile, there is a positive correlation between $T_0$ and
$\beta_T$ as shown in Fig. 8(c). Some studies show negative
correlation between $T_0$ and $\beta_T$ when one uses the
blast-wave model~\cite{22,23,24,25,26,27}. At present, one could
not decide which correlation is correct. In our opinion, for a
given $p_T$ spectrum, $T_0$ and $\beta_T$ are negatively
correlative if one uses the blast-wave model which gives a large
$T_0$ to correspond to a small $\beta_T$. However, for a set of
$p_T$ spectra with varying centralities and energies, the
situation is dissimilar. The present work uses an alternative
method to extract $T_0$ and $\beta_T$ and obtains a positive
correlation.

In addition, one can see some fluctuations in Figs. 3, 4, 5, 6,
and 8. These fluctuations have no particular physics meaning, but
reflect the statistical and/or systematical fluctuations in the
data itself. Although there are fluctuations in the dependences of
parameters on centrality, one can see the general decreasing trend
of parameters with decreasing the centrality. The considered
parameters have similar trend due to their consistent meaning on
the energy deposition which is reflected in terms of excitation
and expansion degree.

Both the mean transverse momentum $\langle p_T\rangle$ and initial
temperature $T_i$ are obtained from $p_T$ spectra. The relation of
$\langle p_T\rangle$ and $T_i$ is certainly positive correlation.
Figure 9 shows this correlation. One can see that $\langle
p_T\rangle$ increases with the increase of $T_i$. This correlation
is natural due to $T_i$ is defined by $\sqrt{\langle
p_T^{2}\rangle/2}$. One can also see that $\sqrt{\langle
p_T^{2}\rangle/2}$ increases with the increase of collision energy
and the size of system, but the later has a very small effect, so
we can neglect it.

From Figs. 4--9 one can see that the considered quantities in the
most peripheral Pb-Pb collisions at 2.76 TeV overlaps the most
central Au-Au collisions with lower energy of 200 GeV, which may
be the indication of formation of similar systems in the most
peripheral collisions at higher energies, and in the most central
collisions at lower energies and it may support the hypothesis of
the effective energy for the particle
production~\cite{48,49,50,51}.

In the considered Au-Au collisions at 200 GeV and Pb-Pb collisions
at 2.76 TeV, the decreasing trend of temperatures and kinetic
freeze-out volume from central to peripheral collisions renders
that more energy deposition and then higher excitation and quicker
expansion in central collisions. Some studies~\cite{9,13,52,53,57}
which use other methods such as the blast-wave model show that the
kinetic freeze-out temperature in central collisions is less than
that in peripheral collisions, though this opposite result can be
explained as longer freeze-out time in central collisions.

Indeed, the kinetic freeze-out temperature and transverse flow
velocity and other quantities are model-dependent. We notice that
the current blast-wave model uses a small or almost zero
transverse flow velocity in peripheral collisions and obtains a
larger kinetic freeze-out temperature in peripheral collisions
comparing to that in central collisions. If we use a large
transverse flow velocity in peripheral collisions, we can obtain a
smaller kinetic freeze-out temperature in peripheral collisions
comparing to that in central collisions~\cite{58}.

In addition, the kinetic freeze-out temperature $T_0$ and
transverse flow velocity $\beta_T$ are also transverse momentum
range dependent. In our opinion, to obtain the parameters as
accurately as possible, we should use the transverse momentum
range as accurately as possible. The transverse momentum range
should not be too narrow or too wide. A too narrow transverse
momentum range will exclude the contributions of some particles
which should be included. A too wide transverse momentum range
will include the contributions of some particles which should be
excluded. In fact, model- and $p_T$-range-independent $T_0$ and
$\beta_T$ are ideal.

Generally, the mean transverse momentum $\langle p_T\rangle$ and
the root-mean-square transverse momentum $\sqrt{\langle
p_T^{2}\rangle}$ are model-independent. Obtaining the initial
temperature by $T_i=\sqrt{\langle p_T^{2}\rangle/2}$ is a suitable
treatment which is regardless of model, though it is from the
color string percolation model~\cite{45,46,47}. It is expected
that $T_0$ and $\beta_T$ are related to $\langle p_T\rangle$,
which results in model-independent $T_0$ and $\beta_T$.

As what we did in our recent work~\cite{59}, let $T_0\equiv
k\langle p_T\rangle/2$ and $\beta_T\equiv (1-k)\langle
p_T\rangle/2\overline{m}$, where $k$ is a parameter which can be
approximately taken to be $0.3-0.01\ln(\sqrt{s_{NN}})$
($\sqrt{s_{NN}}$ is in the units of GeV)~\cite{60}, $1/2$ is used
due to both contributions of projectile and target participants,
and $\overline{m}$ denotes the mean energy of the considered
particles in the source rest frame. If $p_T$-range is wide enough,
$T_0$ and $\beta_T$ are also $p_T$-range-independent.

If we use $T_0$ by the above new definition instead of the
intercept in the linear relation between $T$ and $m_0$, the mean
$T_0$ ($\sim0.10-0.12$ GeV) obtained from Fig. 9 is obviously less
than that in the present work which is too large comparing to
others. Meanwhile, if we use $\beta_T$ by the above new definition
instead of the slope in the linear relation between $\langle
p_T\rangle$ and $\overline{m}$, the mean $\beta_T$
($\sim0.23-0.28$ $c$) obtained from Fig. 9 is less than those in
the present work. Regardless of size, the new definitions of $T_0$
and $\beta_T$ are model-independent.

Before summary and conclusions, we would like to point out that
this paper fits only the transverse momentum spectra measured from
collisions with varying centralities by the two-component standard
distribution. Some centrality dependence of related parameters are
found. In our recent work~\cite{60,61,62,63}, we have fitted the
spectra measured from collisions with varying energies by the
(two-component) standard distribution and/or Tsallis statistics.
Some spectra are from small system size and others are from large
system size. The related parameters are found to depend also on
energy and the larger nucleus in projectile and target nuclei.

In particular, with the increasing energy, the kinetic freeze-out
temperature increases quickly from a few GeV to around 10 GeV and
then slowly or slightly from around 10 GeV to more than 10 TeV.
This implies that around 10 GeV is a special energy at which the
interaction mechanism had changed. In fact, the collision system
undergone from baryon-dominated to meson-dominated final
state~\cite{64}. This implies that the critical energy of phase
transition from hadronic matter to QGP is possibly existent at
around 10 GeV.

The dependence on the larger nucleus in projectile and target
nuclei is consistent to the dependence on centrality. This implies
possibly that there is a critical size from small to large system,
and from peripheral to central collisions. The data measured by
the NA61/SHINE Collaboration ~\cite{65} show that the nucleon
number in projectile or target nucleus on the onset of
deconfinement is $\approx10$. Meanwhile, the energy on the onset
of deconfinement is $\approx10$ GeV. This double 10 signature is
very interesting and should be studied further by various models
and methods in future.

We have studied three types of temperatures, namely the effective
temperature, initial temperature, and kinetic freeze-out
temperature, in this paper. Although the three types of
temperatures are extracted from the transverse momentum spectra,
they have different physics meanings. The effective temperature is
obtained directly from the fit function, which describes together
the degree of the thermal motion and flow effect at the stage of
kinetic freeze-out. In the case of excluding the contribution of
flow effect from the effective temperature, we expect to obtain
the kinetic freeze-out temperature which describes only the
thermal motion. The initial temperature in this paper is quoted
directly from the color string percolation model~\cite{45,46,47},
which is expected to describe the excitation degree of initial
state as what we did in our recent work~\cite{65a}.

It is regretful that the chemical freeze-out temperature is not
discussed in this paper, though it has wider applications and
discussions in literature~\cite{66,67,68}. The chemical freeze-out
temperature describes the excitation degree of collision system at
the stage of chemical freeze-out. Generally, the chemical
freeze-out temperature can be obtained from the ratio of particle
yields, and can be used to map the phase diagram with the chemical
potential. In the extensive statistics and/or axiomatic/generic
non-extensive statistics~\cite{66,67,68}, one may discuss the
chemical and/or kinetic freeze-out parameters systematically.
\\
\\

{\section{Summary and conclusions}}

The main observations and conclusions are summarized here.

(a) The transverse momentum spectra of $\pi^-$, $K^-$ and $\bar p$
at mid-(pseudo)rapidity produced in different centrality events in
Au-Au collisions at 200 GeV and Pb-Pb collisions at 2.76 TeV have
been analyzed. The experimental data measured by the PHENIX and
ALICE Collaborations are fitted by the two-component standard
distribution in which the temperature concept is quite close to
the ideal gas model.

(b) The effective temperature, initial temperature, kinetic
freeze-out temperature, transverse flow velocity and mean
transverse momentum increase with the increase of event centrality
from peripheral to central collisions, which results in higher
excitation degree and quicker expansion velocity in central
collisions. The kinetic freeze-out volume increases with the
increase of event centrality from peripheral to central collisions
due to more number of participant nucleons taking part in central
collisions.

(c) The mass dependent differential effective temperature, initial
temperature, kinetic freeze-out volume and mean transverse
momentum are observed. The kinetic freeze-out temperature and
transverse flow velocity extracted in this paper does not show
mass dependent differential scenario due to the reason of
methodology. Many quantities are model or method dependent.

(d) The formation of similar system is possible in the most
peripheral nucleus-nucleus collisions at high energy and in the
most central nucleus-nucleus collisions at low energy. This
observation confirms the hypothesis of the effective energy for
the particle production. There are many similarities in high
energy collisions.
\\
\\
{\bf Data availability}

The data used to support the findings of this study are included
within the article and are cited at relevant places within the
text as references.
\\
\\
{\bf Ethical Approval}

The authors declare that they are in compliance with ethical
standards regarding the content of this paper.
\\
\\
{\bf Disclosure}

The funding agencies have no role in the design of the study; in
the collection, analysis, or interpretation of the data; in the
writing of the manuscript; or in the decision to publish the
results.
\\
\\
{\bf Conflict of Interest}

The authors declare that there are no conflicts of interest
regarding the publication of this paper.
\\
\\
{\bf Acknowledgments}

This work was supported by the National Natural Science Foundation
of China under Grant Nos. 11575103 and 11947418, the Chinese
Government Scholarship (China Scholarship Council), the Scientific
and Technological Innovation Programs of Higher Education
Institutions in Shanxi (STIP) under Grant No. 201802017, the
Shanxi Provincial Natural Science Foundation under Grant No.
201901D111043, and the Fund for Shanxi ``1331 Project" Key
Subjects Construction.
\\
\\

\end{multicols}
\end{document}